\documentclass[twocolumn,showpacs,amsmath,amssymb]{revtex4-1}
\usepackage{epsfig}
\usepackage{amsmath}
\usepackage{amssymb}

\DeclareMathAlphabet{\bi}{OML}{cmm}{b}{it}

\begin{document}

\title{ Off-resonant polarized  light-controlled %exchange of states  and sign reversals in thermoelectric 
thermoelectric transport \\in ultrathin topological insulators } %under %polarized 
%off-resonant light}  %polarization-induced 
%exchange of states  and sign reversals}  
%\title{ Thermoelectric transport in ultrathin topological
%insulators under polarized off-resonant light:  %polarization-induced 
%exchange of states  and sign reversals}  %the magnetization and Nernst  conductivity}
\author{M. Tahir$^{\dag}$ and P. Vasilopoulos$^{\star}$}
\affiliation{Concordia University, Department of Physics, 7141 Sherbrooke Ouest, Montr\'{e}al, Qu\'{e}bec, Canada, H4B 1R6}

\begin{abstract}
We study  thermoelectric transport  in ultrathin topological insulators under the application of circularly
polarized \textit{off-resonant light} of frequency $\Omega$ and amplitude $A$. We derive analytical expressions for the band structure, orbital magnetization $M_{orb}$,  
and the thermal ($\kappa_{xy}$) and Nernst ($\alpha_{xy}$) conductivities. %Chang
Reversing the  light polarization from right  to left leads to an exchange   
of the conduction and valence bands of the symmetric and antisymmetric surface states and to 
a sign change in $M_{orb}$, $\alpha_{xy}$, and  $\kappa_{xy}$. %and the Nernst and thermal conductivities 
%change sign. 
Varying the sample thickness or $A/\Omega$  leads to a strong %large 
enhancement of $M_{orb}$ and $\alpha_{xy}$. %the orbital magnetic moment 
%and a sign reversal of the Nernst and thermal conductivities, and of the orbital magnetization. 
These  effects,  accessible to experiments,  
open the possibility for selective, state-exchanged excitations under light and the conversion of heat to electric energy. 
\end{abstract}
% \vspace*{-0.8cm}
\maketitle
 
 \vspace*{-0.6cm}
\textit{Introduction.}
Topological insulators (TIs) are  insulators in the bulk but possess  gapless surface
states \cite{MZ%JE,XLQ
}. Due to their potential applications  a wide variety
of TIs has been found to be three dimensional and among them Bi$_{2}$Se$%
_{3}$ and Bi$_{2}$Te$_{3}$ are demonstrated as prototypes with  single
Dirac-cone surface states due to the strong spin-orbit interaction (SOI) \cite%
{HZ,YZ,SC,HL,XZ}. By reducing their thickness 
to 6 nm or less, a finite hybridization occurs of their top and bottom surface states \cite%
{YZ,XZJ} 
and has been realized in transistors \cite{SCN}.  Ultrathin  TIs 
are %expected to be 
promising materials for high performance optoelectronic devices, such as
photodetectors \cite{XZJ} and transparent electrodes \cite{HP},  and excellent thermoelectric materials 
\cite{PG,TY}.
 
Berry-curvature mediated thermoelectric effects,  generated by a temperature gradient \cite{KU},
have been proposed for  two-dimensional (2D) systems and explained  related
experiments very well \cite{DX}. Among their properties, the orbital magnetic
moment and corresponding orbital magnetization \cite{DXM,SY}, and the thermal and
Nernst conductivities \cite{PW,YM,KL,AA} have attracted considerable attention. 
%Moreover, 
Also, Berry-curvature
mediated transverse %magnetic 
heat transport on the surface of TIs attached to
a ferromagnet has  been demonstrated \cite{TY} despite the complicated
nature of the experiments.   Of particular interest  
is the control of thermoelectric effects through the surface
states of TIs under circularly polarized \textit{off-resonant light} \cite%
{TK}.

%\textbf{
In recent years light-induced quantum effects have generated a strong interest, in particular quantum phase transitions in Floquet TIs driven by external time-periodic perturbations \cite{NHL}.
%have stimulated great interests in this direction .in particular, for time-dependent 
For such systems it is convenient to use Floquet theory proposed \cite{TK} recently for periodically driven %nonequilibrium  systems (
 graphene and TIs. In the appropriate frequency regime the off-resonant light cannot generate real photon absorption or emission due to energy conservation.  Accordingly, it does not directly 
 excite electrons but instead %effectively 
  modifies the electron band structure through second-order virtual-photon absorption processes.
 % . The modification of the system Hamiltonian 
 % can be understood by second-order virtual photon processes. 
%  (a photon is first absorbed/emitted and then emitted/absorbed). 
  When averaged over time, these processes result in an effective static alteration of the band gap of the system.
  
     Floquet bands were first realized in photonic crystals \cite{MCR} and have been verified by recent experiments on surfaces of TIs \cite{YH,HZJ}.
%
%\textbf{
%The effect of this  light, first predicted 
%\cite{TK} for graphene and the surface states of TIs,  has been
%confirmed experimentally for Tis \cite{YH,HZJ}.
%...we need to delete this} The development of
%new experimental probes \cite{YH,HZJ} makes it possible to access,
%through circularly polarized light,  new and rich %and versatile aspects of  
%topological phases  
%of TIs. 
However, nontrivial  phase transitions, induced by \textit{off-resonant light} on the
surface states of  ultrathin TIs, and the effect of this light on transport properties
is an open question
as it is not yet studied and is different than many optical effects in TI films %studied in many works 
\cite{OE}.
 In this work we partly answer this question by evaluating  
 the band structure, %orbital magnetic moment,
  orbital  magnetization, and the thermal and Nernst conductivities of such   TIs. Reversing the polarization of this light  leads to an exchange  
of the conduction and valence bands of the symmetric and antisymmetric surface states  
and  a tunable  band gap.  The details are as follows.

 \textit{Basic Formalism}. 
We consider surface states of  ultrathin
TIs in the (x,y) plane in the presence of circularly polarized
light and hybridization between the top and bottom surface states. We first extend the
2D Dirac type Hamiltonian \cite{XZJ} by including a time-periodic field 
\cite{TK} as
\begin{equation}
H(t)=v_{F}(\sigma _{x}\Pi _{y}(t)-\sigma _{y}\Pi _{x}(t))+s\Delta _{h}\sigma
_{z}.  \label{1}
\end{equation}
Here $s=\pm 1$ for symmetric and antisymmetric combinations of the two
surface states, ($\sigma _{x}$, $\sigma _{y}$, $\sigma _{z}$) are the Pauli
matrices, $v_{F}$  the Fermi velocity, % of Dirac fermions, 
and $\Delta _{h}$  the hybridization energy between the top and bottom surface states that, depending on the thickness, varies from 41 meV to 250 meV
\cite{YZ,XZJ}. %\textbf{
For simplicity we
disregard   higher order terms in $k$  since their contribution  is very small and doesn't affect the major physics discussed below \cite{HZL,JLT,CXL}. Further,  $\mathbf{\Pi (t)=P}+e\mathbf{%
A(t)}$ is the  canonical momentum with vector potential $\mathbf{A}(t)=(lA\sin (\Omega t),A\cos \Omega t)$, 
$\mathbf{E}(t)=-\partial \mathbf{A}(t)/\partial t$ the %time-periodic 
electric field with amplitude $E_{0}$,  $\Omega $ 
the light's frequency, %of the light, 
and $A=E_{0}/\Omega $. The gauge potential is periodic in time,
 $A(t+T)=A(t)$, with period $T=2\pi /\Omega $ and $l$ = $ 1 (-1)$,
stands for the right (left) %circular 
polarization of the
light. Equation (1) can be treated by the Floquet method with the
aid of an  effective static Hamiltonian %constructed appropriately
 \cite{TK}
and leads to results  %in excellent 
that agree well with  experiments \cite%
{YH,HZJ}. This formalism has been successfully applied to graphene \cite%
{AG} and disordered TIs \cite{PT}. %Following this method
% in the limit of 
For  high frequencies $\Omega \gg ev_{F}E_{0}/\hslash
\Omega $ and low intensities ($ev_{F}A\ll \hslash \Omega $) 
%this method 
it gives
%of light,
%we obtain %
the effective static Hamiltonian
\begin{equation}
H_{l}^{s}=v_{F}(\sigma _{x}p_{y}-\sigma _{y}p_{x})+s\Delta _{h}\sigma
_{z}+l\Delta _{\Omega }\sigma _{z},  \label{2}
\end{equation}
where 
$\Delta _{\Omega }=e^{2}v_{F}^{2}\hslash ^{2}A^{2}/\hslash
^{3}\Omega ^{3}$ is the mass term induced by the off-resonant light. It breaks the time-reversal symmetry and its values are in the range of 100 meV \cite{YH,HZJ}.
The diagonalization of  Eq. (2) gives the eigenvalues  
\begin{equation}
E_{l}^{s\lambda }=\lambda \left[ \varepsilon _{k}^{2}+\delta _{sl}^{2}% 
\right] ^{1/2} \label{3}
\end{equation}
and the corresponding eigenfunctions 
\begin{equation}
\Psi _{l}^{s,\lambda }=(1/\sqrt{S})\exp [i\mathbf{k}\cdot \mathbf{r}]\binom{i\varepsilon _{k}e^{-i\varphi
}/Y}{X/Y}.  \label{4}
\end{equation}
Here $S=L_{x}L_{y}$, $\lambda=\pm 1$, $\varepsilon _{k}=v_{F}\hslash k$, 
$\delta_{sl}=l\Delta _{\Omega }+s\Delta _{h}$, $Y^2=\varepsilon _{k}^{2}+X^{2} %]^{1/2}
$, $X=E_{l}^{s,\lambda }-\delta _{s,l}$, $\tan \varphi = 
k_{y}/k_{x}$, and $k^2=k_{x}^{2}+k_{y}^{2}$. 
We show the eigenvalues given by Eq. (3)  in Fig.\ 1 for the symmetric
(solid curves) and antisymmetric (dashed curves) states. We fixed the
hybridization energy to $35$ meV corresponding to 4 quintuple layers \cite{YZ}, 
$v_{F}=0.5\times 10^{6}$ m/s, $a$ = 4.14 \AA\ , $\hslash \Omega =8J=8$ eV
($J$ is the nearest neighbor hopping amplitude), $\Delta _{\Omega }=20$
meV ($ev_{F}A=0.4$ eV) \cite{TK}. We find a well resolved  gap between
the valence and conduction bands and notice that the surfaces are
nondegenerate if $\Delta _{\Omega }\neq 0$ and $\Delta _{h}\neq 0$. Here we
vary the amplitude of the circularly polarized {\it off-resonant light} such that
$\Delta _{\Omega }=0$ meV, $20$ meV, $35$ meV ($ev_{F}A=0.53$ eV), and $100$ meV ($ev_{F}A=0.9$
eV), which can be achieved by existing experimental techniques \cite%
{YH,HZJ}. As $\Delta _{\Omega }$ at $k=0$ changes  sign when
the light  polarization  is reversed. The realization of
this reversed state of the system upon changing the light polarization from right to left  is an entirely new
phenomenon; it is made clear  upon contrasting Fig. 2 with Fig. 1.
\begin{figure}[ht]
\begin{center}
 \vspace*{-0.3cm}
\includegraphics[width=0.48\columnwidth,height=0.4\columnwidth %clip
]{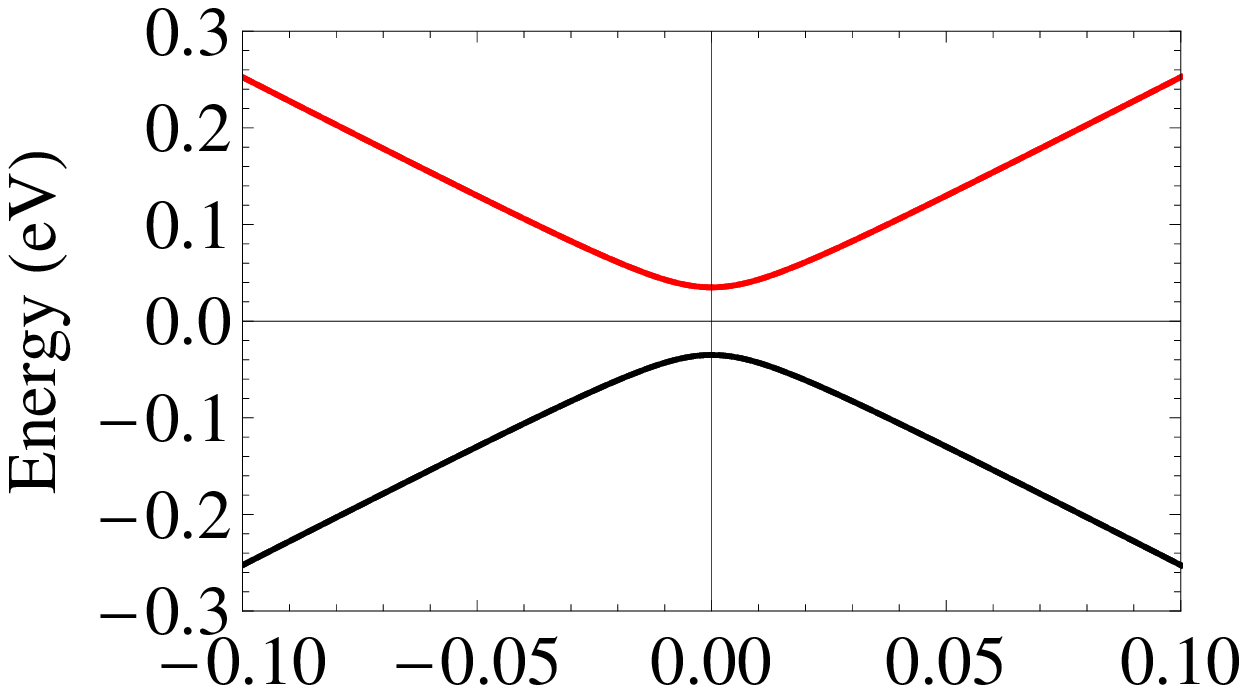}\vspace{-0.15 cm} %
\includegraphics[width=0.48\columnwidth,height=0.4\columnwidth %clip
]{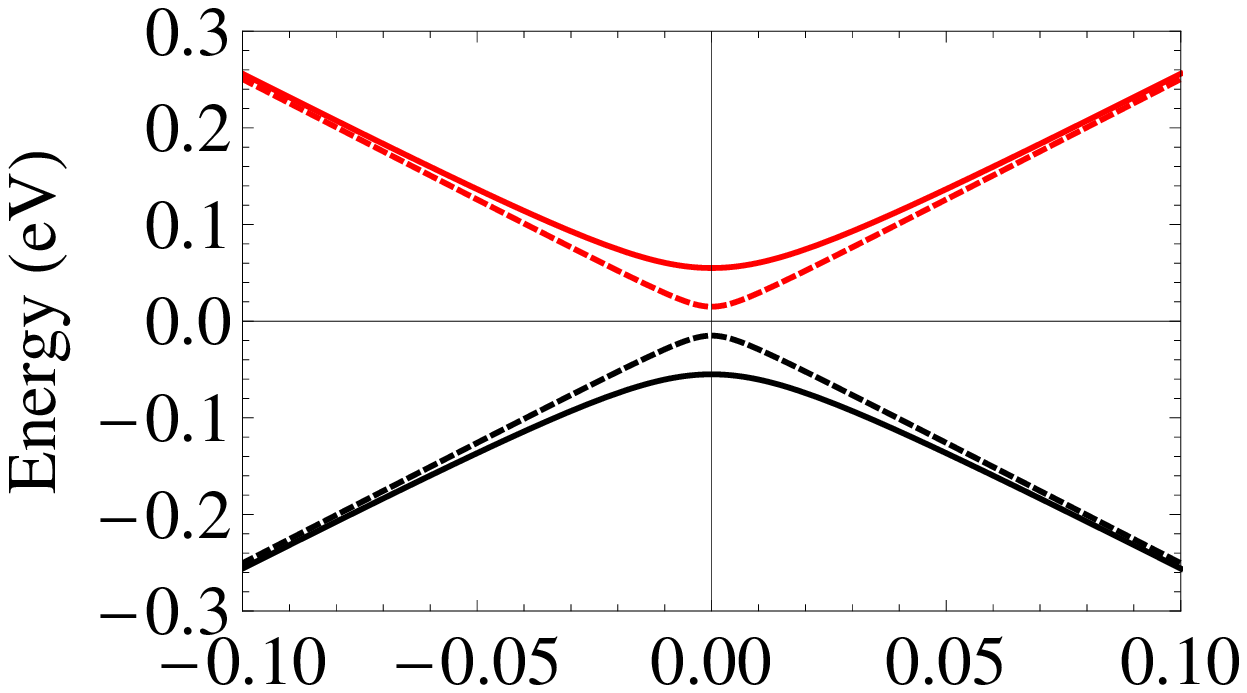}\vspace{-0.15 cm} %
\includegraphics[width=0.48\columnwidth,height=0.4\columnwidth %clip
]{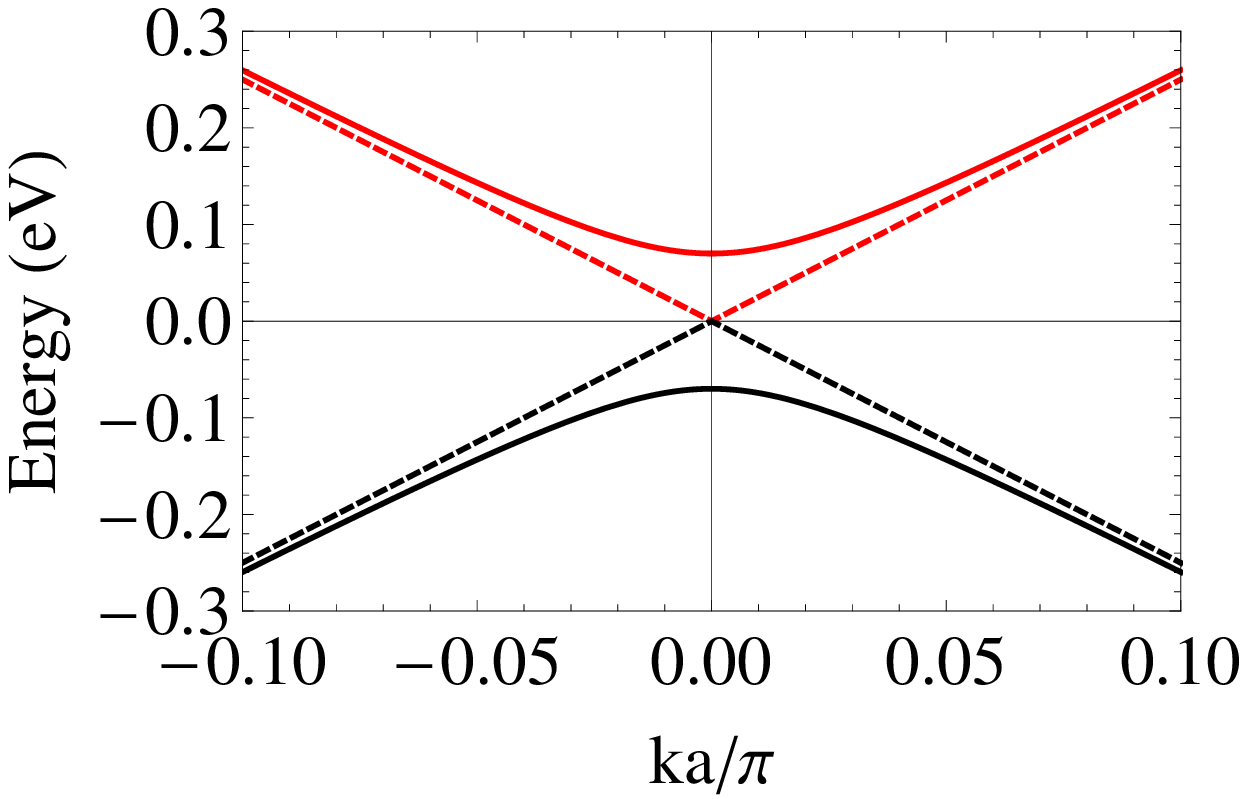} %
\includegraphics[width=0.48\columnwidth,height=0.4\columnwidth %clip
]{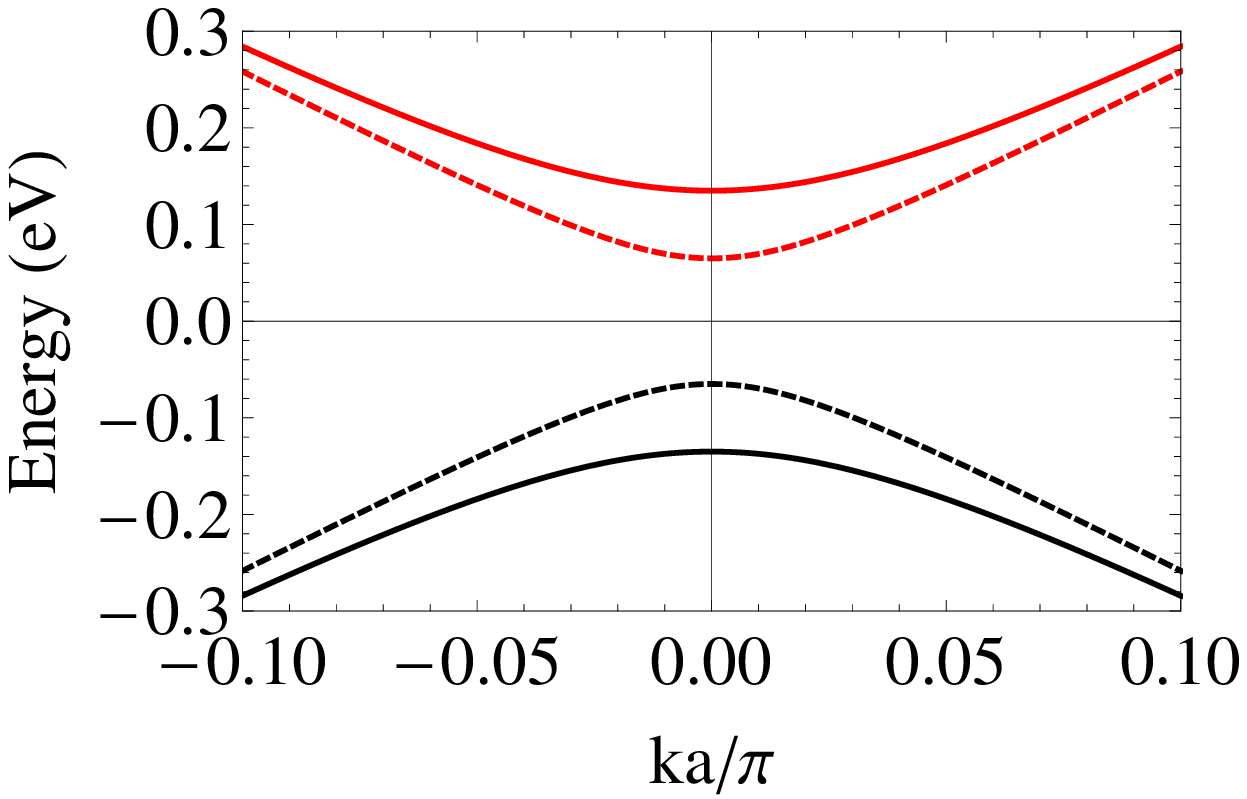}
\end{center}
\vspace{-0.62cm}
\caption{Band structure  for
right-% circularly 
polarized light ($l=+1$). Here $\Delta_{h}=35$ meV and $%
\Delta_{\Omega} = 0$ meV (top left), $20$ meV (top right), $35$ meV (bottom
left), and $100$ meV (bottom right). The solid (dashed) curves correspond to the
symmetric (antisymmetric) surface states. Red and black
colours represent the conduction and valence bands, respectively.}
\label{fig:1}
\end{figure}

\textit{Orbital magnetization}. The orbital
magnetization of Bloch electrons has been a very attractive problem since the
prediction and observation of its dependence on the Berry-curvature  \cite%
{DX,DXM}. To study  thermoelectric transport properties we must include  
their temperature dependence. We obtain the equilibrium
magnetization density from the %total 
free energy $F$ in
a weak magnetic field $B$ ($B$ only couples to
the orbital motion of electrons but does not contribute to the Zeeman energy \cite{DXM}) written as 
\begin{equation}
F=-\frac{1}{\beta }\sum_{s\lambda \mathbf{k}} \ln \Big(1+e^{\beta
(\mu-E_{l\mathbf{k}}^{s\lambda })}\Big).  \label{5}
\end{equation}
Here $\beta =1/k_{B}T, k_{B}$   is the Boltzmann constant, 
 $T$ the temperature, and $\mu $  the chemical potential. Further, the electron energy $E_{l\mathbf{k}}^{s\lambda }=E_{l}^{s\lambda }-%
\mathbf{m}\mathbf{(k)}\cdot\mathbf{B}$ includes a correction due to
the orbital magnetic moment $\mathbf{m}\mathbf{(k)}=(-ie/
2\hslash)\left\langle \nabla _{k}\Psi _{l}^{s\lambda }\left\vert \times
\lbrack \hat{H}-E_{l}^{s\lambda }]\nabla _{k}\right\vert \Psi
_{l}^{s\lambda }\right\rangle$.  
%This moment originates from the self-rotation of the electron wave packet around its center of mass \cite{DXM}.

To evaluate  the sum over $\mathbf{k}$ in Eq. (5) %and (6) 
we convert it to an integral and use the prescription % \cite{DXM} 
$\sum_{\mathbf{k}}\to(1/2\pi )^{2})\int d^{2}\mathbf{k}(1+e\Omega
(\mathbf{k})\cdot\mathbf{B}/\hslash )$, where $\Omega (\mathbf{k})=\mathbf{\nabla }_{k}\times \left\langle
\Psi _{l}^{s\lambda }\left\vert i\mathbf{\nabla }_{k}\right\vert \Psi_{l}^{s\lambda }\right\rangle $ is the Berry-curvature, see Ref.  \onlinecite{DXM} for  a detailed justification. %\textbf{
The orbital magnetization $M_{orb}=M_{c}+M_{\Omega}$ is
given  by $M_{orb}=-(\partial F/\partial B)_{\mu ,T}$. $M_{c}$ is the conventional term  and %$M_{c}$ %in terms of orbital magnetic moment, 
%we have the extra term 
$M_{\Omega}$ the additional term due to the  Berry curvature. 
It originates from the self-rotation of the electron wave packet around its center of mass \cite{DXM}.
% effect.} 
The results for  $M_{c}$ and  
$M_{\Omega}$ are
 %
%\vspace*{-1.2cm}
\begin{eqnarray}%uation} 
%\nonumber
M_c&=&(1/2\pi )^{2}\sum_{s\lambda}\int m(\mathbf{k}) f(\mathbf{k})\,d^2k, %\mathbf{k} 
%\,(1+e\Omega
%(\mathbf{k})\cdot\mathbf{B}/\hslash )
\\*
&\hspace*{-.6cm}M_{\Omega}=&%\big[f(\mathbf{k})\mathbf{m}_{orb}\mathbf{(k)}+%\frac{1}{
(e/2\pi\beta h) \sum_{s\lambda}\int \Omega
(\mathbf{k})%\cdot\mathbf{B}%(e/\hslash )
\ln \Big(1+e^{\beta (\mu-E_{l\mathbf{k}}^{s\lambda })}\Big)%\big]
\,d^2k,  \label{6}
\end{eqnarray} 
with $f(\mathbf{k})$ the Fermi function. %Using 
Equations (3) and (4) give %we obtain  
\begin{equation}
\Omega (\mathbf{k})=%\frac{
(\hslash ^{2}v_{F}^{2}/2) %}{2} %\frac{
\,\delta _{sl}\big/%}{
\big(
\varepsilon _{k}^{2}+\delta _{sl}^{2}\big) ^{3/2}.  \label{7}
\end{equation}
%
%and  $m(\mathbf{k})=(e/\hslash)%
%E_{l}^{s\lambda }\Omega (\mathbf{k})$ \textbf{is obtained using procedure given in Ref. \cite{DWQ}.} 
For finite $\Delta _{h}$ or $\Delta_{\Omega }$ the  moment  $m(\mathbf{k})$ has a peak at $k=0$. For  
$\Delta _{\Omega }=0$  and $\Delta _{h}$ = 35 meV we obtain $m(\mathbf{k})= 
25$ Bohr magnetons. This value may be changed  by varying the light intensity   $\Delta_{\Omega }$. %\textbf{It is important to 
%Notice that for %in the limit of very low temperatures 
%($T\to 0$) and $\mu>E_{l\mathbf{k}}^{s\lambda }$ we have 
% $\ln(...)\approx \beta(\mu-E_{l\mathbf{k}}^{s\lambda }) $ and  the term $E_{l\mathbf{k}}^{s\lambda } $ cancels $M_{c}$. %accordingly 
 Then only %we are$M_{\Omega}$ and we are 
the  purely Berry-curvature-mediated orbital magnetization survives. 
%This means that the two parts make opposite contributions to the orbital magnetization and  implies that they carry opposite circulating currents. 
For a qualitative analysis we obtain 
$M_{orb}$ at $T=0$ and %the orbital magnetization at  zero temperature 
 $\mu$ in the conduction band, as
\begin{equation}
M_{orb}=(e\mu/2h)
\sum_{s} \Big[1-\delta _{sl}/\big(\varepsilon_{k_{F}}^{2}+\delta _{sl}^{2}\big) ^{1/2}\Big] . 
\label{8}
\end{equation}
 %
%where $\mu\approx E_{F}^{'}=\big[ \varepsilon_{k_{F}}^{2}+\delta _{sl}^{2}\big] ^{1/2}$. 
%
%We need to explain the confusion of $E_F$ in above eq. 
%
\begin{figure}[ht]
\begin{center}
\vspace{-0.4cm}
\includegraphics[width=0.48\columnwidth,height=0.4\columnwidth %clip
]{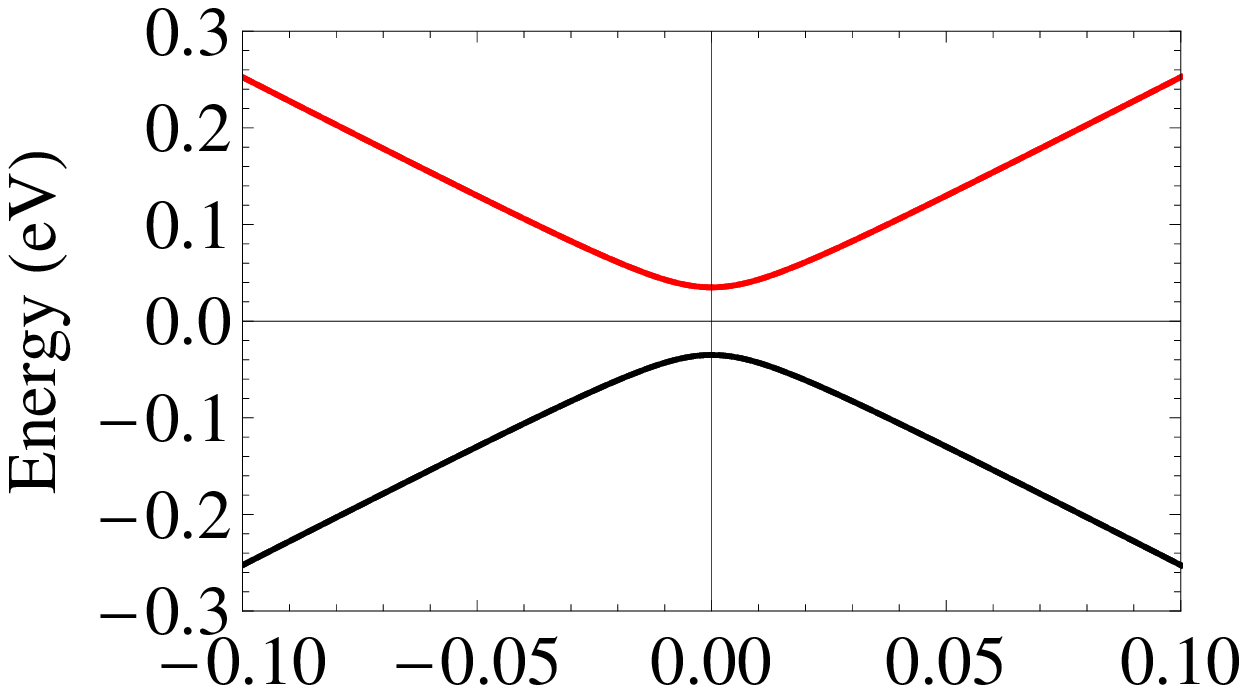}\vspace{-0.15 cm} %
\includegraphics[width=0.48\columnwidth,height=0.4\columnwidth %clip
]{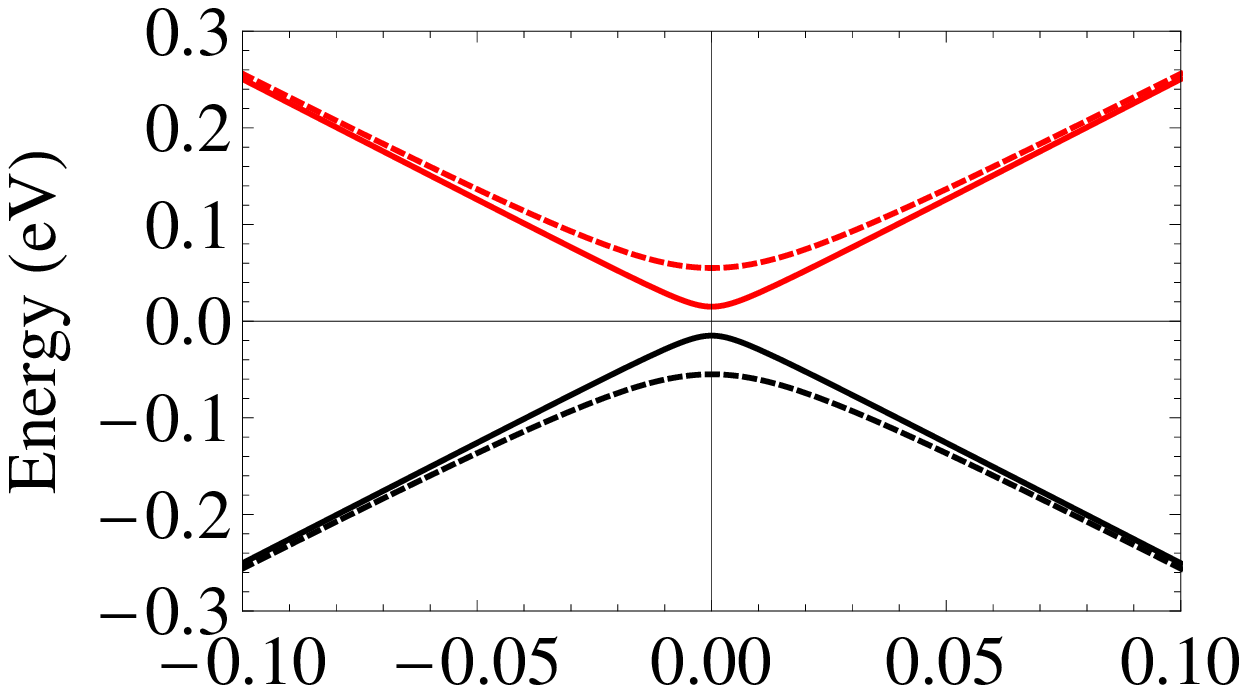}\vspace{-0.15 cm} %
\includegraphics[width=0.48\columnwidth,height=0.4\columnwidth %clip
]{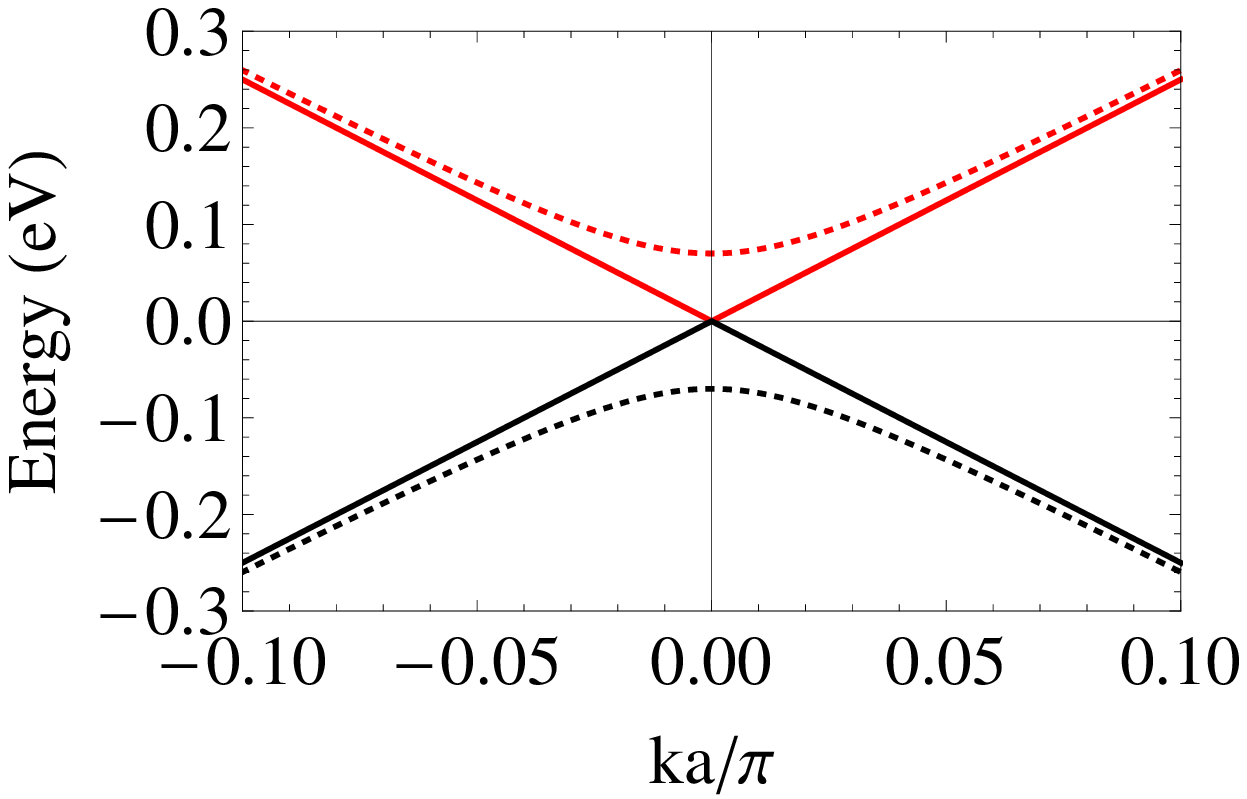} %
\includegraphics[width=0.48\columnwidth,height=0.4\columnwidth %clip
]{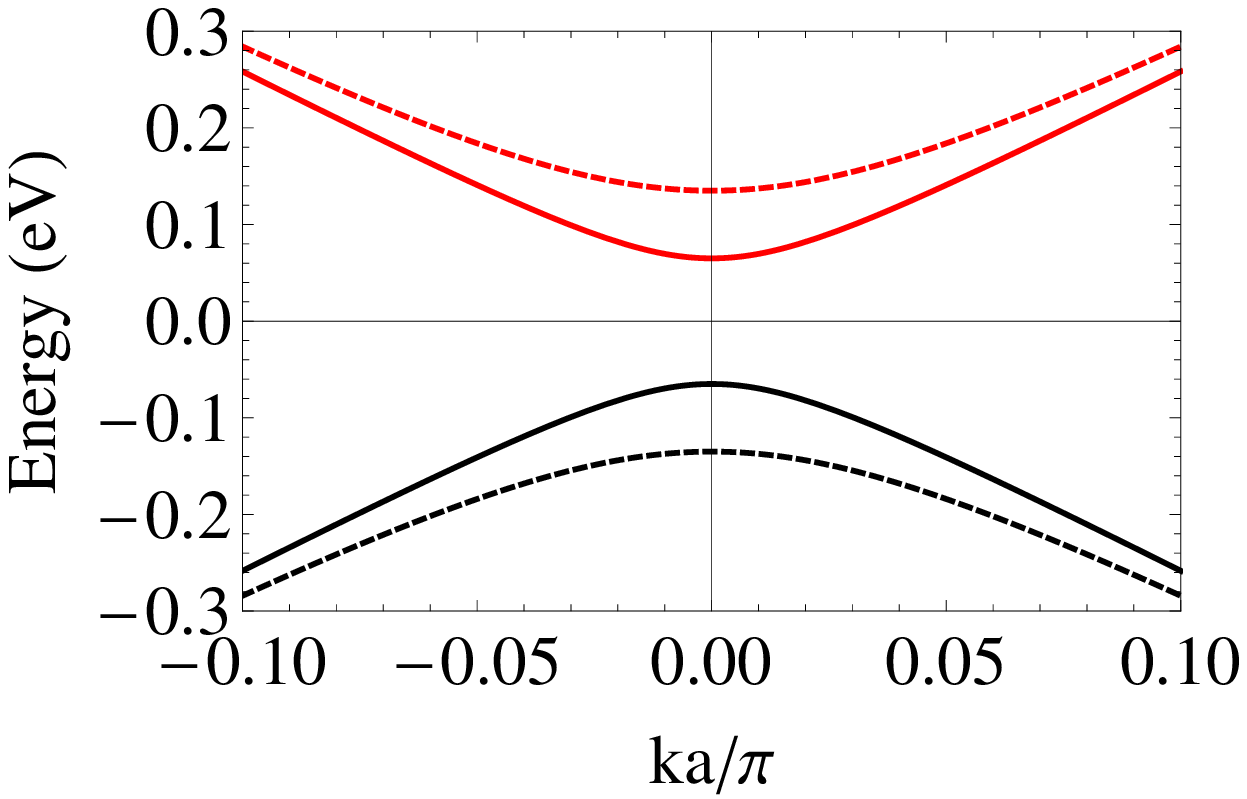}
\end{center}
\vspace{-0.6cm}
\caption{As in Fig. 1 for
left-% circularly 
polarized light ($l=-1$). All other parameters are the same
as in Fig. 1}
\label{fig:2}
\end{figure}
%
%\textbf{ 
 We show $M_{orb}$,  obtained numerically from Eqs. (6) and (7), in Fig. 3. Its magnitude  can be
enhanced by tuning the band gap upon 
varying $\Delta _{\Omega }$ and/or $\mu$.  As a
reference, for $\mu=100$ meV and a typical layer thickness $d$ of
2 nm, Eq. (9) gives  $M_{orb}=0.004$ T. %esla . 
The value of $M_{c}$, evaluated in the manner of Ref. \onlinecite{DWQ}, is one order of magnitude smaller than the $M_{orb}$.
Magnetization values as small as $10$ mT have been recently measured in a Cr-doped Bi$_2$(Se, Te)$_3$ TI \cite{col}. 
The weak cusps at $\mu=\pm 45$ meV and $\pm 115$ meV are due to the  
gap induced by $\Delta _{\Omega }$ and $\Delta
_{h}$; they are  washed out as we raise  the temperature to 150 K (dotted).
Notice though that the temperature dependence is weak. 
The orbital magnetization induced by \textit{off-resonant light} can be  
 distinguished from other   sources of magnetization that don't
depend on the polarization of light,  e.g., from the one induced by spin-orbit coupling \cite{gru}. % 
We can clearly see from Eq. (9) that $M_{orb}$ %it  
changes sign due to $l\Delta _{\Omega }$, for  $\Delta _{\Omega }>\Delta _{h}$, when we reverse   the light's polarization ($l=\pm 1)$.  
This could be observed in  experiments
similar to those on magnetization  
\cite{SY} or by the Faraday-Kerr effect 
\cite{YK}. Moreover, very recently magnetization signatures of the
off-resonant light effects  on graphene have been simulated and helical edge
states have been reported \cite{JP}.
\begin{figure}[ht]
\begin{center}
\vspace{-0.3cm}
\includegraphics[width=0.9\columnwidth,height=0.4\columnwidth %clip
]{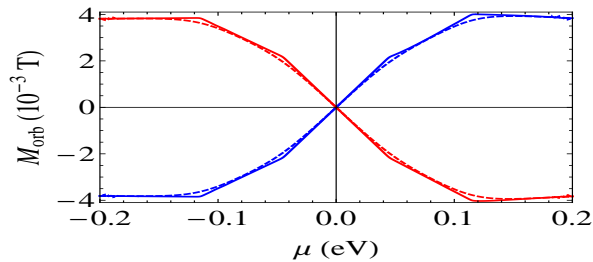}
\end{center}
\vspace{-0.55cm}
\caption{$M_{orb}$  divided by the layer thickness (2 nm) vs 
$\mu$ for $T$ = 10 K (solid) and 150 K (dotted). Here $\Delta_{h}=35$ meV and $%
\Delta_{\Omega}$ = 80 meV. Blue and red curves correspond to  right- ($l=+1$%
) and left- ($l=-1$) %circularly 
polarized light, respectively.}
\label{fig:3}
\end{figure}

{\it Thermal and Nernst conductivities}. The orbital magnetization
%obtained in Eqs. (6) and (7) 
contains a conventional contribution, Eq. (6),
and a Berry-curvature-mediated  one, Eq. (7). The
relation between it and the Nernst conductivity,  
demonstrated in Refs. \onlinecite{DX,DXM}, shows that the conventional 
part does not contribute to the transport current whereas the
Berry-curvature term  directly does and modifies the
intrinsic Hall current, obtained by an integral with respect to momentum, of the
Berry curvature over the 2D Brillouin zone.
 %(a torus) with periodic boundary conditions. 
 The difference between the Hall  $\sigma _{xy}$ and 
Nernst $\alpha _{xy}$ conductivities is that $\alpha _{xy}$ is determined not only by the
Berry curvature but also by entropy generation around the Fermi surface.
Therefore, $\alpha _{xy}$ is sensitive to changes of the Fermi
energy and  temperature.  The heat current under
a weak electric field $\mathbf{E}$\ and a thermal gradient $\nabla T$ is given  
by $\mathbf{J}^{Q}=T%\boldsymbol{
\alpha\cdot\mathbf{E}-%\boldsymbol{
\kappa\cdot\boldsymbol{\nabla }T$. In this case the intrinsic Hall current is $j_{x}=-\alpha _{xy}\nabla _{y}T$ \cite%
{TY,DXM}. With $ \Omega \equiv\Omega (\mathbf{k})$ the component $\alpha _{xy}$ is given by %reads%is obtained as %given by  
\begin{equation}
%\nonumber
\hspace*{-0.15cm}\alpha _{xy}=c%\frac{e}{2\pi h%\hslash T}
\sum_{\mathbf{s\lambda }}\int %\frac{
%d^{2}k%
%\mathbf{k}}{4\pi^{2}}
\Omega %(\mathbf{k})
\big[ E_{l}^{s\lambda }(%
\mathbf{k})f^{s\lambda}_{l\mathbf{k}} 
\\*
+k_{B}T\ln (1-f^{s\lambda}_{l\mathbf{k}}) 
\big]d^{2}k, \label{9}
\end{equation}
with %$ \Omega \equiv\Omega (\mathbf{k})$, 
$c=e/2\pi hT$,   $f^{s\lambda}_{l\mathbf{k}}= f(E_{l}^{s\lambda}(\mathbf{k} ))$  the Fermi-Dirac  function, and  $\mu$ the chemical potential.
Recent experiments on graphene   \cite{JG} agree %very 
well with Eq. (10). The component $\kappa _{xy}$ of the %corresponding 
thermal conductivity tensor $\kappa$ reads %is given by 
\begin{eqnarray}
%\nonumber
\hspace*{-0.6cm}
\kappa _{xy}=b 
\nonumber
\sum_{\mathbf{s\lambda }}\int d^{2}k\, 
\Omega \Big[ 
\,
\beta ^{2}(E_{l}^{s\lambda })^{2}f^{s\lambda}_{l\mathbf{k}} -2\text{Li}_{2}(1-f^{s\lambda}_{l\mathbf{k}})
\\*
%\ \\ 
+ \pi ^{2}/3-\log^2 \big(1+e^{-\beta E_{l}^{s\lambda }}\big)%]^{2} 
\Big],  
\label{10}
\end{eqnarray}
where $b=ek_{B}/4\pi^{2}\beta h$; Li$_{2}(x)$ is the polylogarithm function. Equations (10) and (11) can be
simplified in the limit of low temperatures  using the Mott relations \cite%
{TY,DXM}, $\alpha _{xy}=-(\pi ^{2}k_{B}^{2}T/3e)(d\sigma _{xy}/d\mu)=-(e/k_{B})d\kappa _{xy}/d\mu$ and $\kappa _{xy}=(\pi^{2}k_{B}^{2}T/3e^{2})\sigma _{xy}$ with $\sigma _{xy}$ given by
\begin{equation}
\sigma _{xy}=\frac{e^{2}}{2\pi h} \sum_{s} \int %\frac{d^{2}\mathbf{k}}{(2\pi )^{2}}%
\Omega\big( f^{s1}_{l\mathbf{k}}-f^{s,-1}_{l\mathbf{k}}\big)\,  d^{2}k %(\mathbf{k}). 
 \label{11}
\end{equation}
For $T=0$ and $\mu$ in the band gap Eq.~(12) gives  $\sigma _{xy}^{0}$, %\textbf{ 
the Hall conductivity in the gap,  as $\sigma _{xy}^{0}=-(e^{2}/2h) sgn(\delta _{sl})$. Here it is interesting to note that for $\Delta _{\Omega }<\Delta _{h}$ the insulating state is trivial whereas for $\Delta _{\Omega }>\Delta _{h}$ the state is topological nontrivial; a   topological phase transition occurs at $\Delta _{\Omega }=\Delta _{h}$. Such transitions were also reported in previous studies %demonstrated 
without off-resonant light \cite{BSJ,HZL,JLT,CXL}.
 For $\mu$
 in the conduction band  we have $\sigma _{xy}\equiv\sigma _{xy}^{c}$ with
\begin{equation}
\sigma _{xy}^{c}=(e^{2}/2h)\sum_{s} %[1-(\delta _{sl}/E_F^{'})]. 
\Big[1-\delta _{sl}/\big( \varepsilon_{k_{F}}^{2}+\delta _{sl}^{2}\big) ^{1/2}\Big] 
\label{12}
\end{equation}
Notice that due to $\delta _{sl}=l\Delta _{\Omega }+s\Delta _{h}$ the  sign  of $\sigma _{xy}^{c}$ can be 
reversed, for $\Delta _{\Omega }>\Delta _{h}$, upon reversing the light polarization ($l\to -l$).
Similar results can be obtained when the chemical potential $\mu$ 
%E_{F}^{'}=\big[ \varepsilon_{k_{F}}^{2}+\delta _{sl}^{2}\big] ^{1/2}$ 
is in the valence
band due to symmetry. For a qualitative analysis we use Eq. (13) and obtain 
$\alpha _{xy}$, at very low temperatures,  as
\begin{equation}
\alpha _{xy}=-(\pi ^{2}\,ek_{B}^{2}T/6h)\sum_{s} %\,\delta _{sl}/(E_{F}^{'})^2    \label{13}
\delta _{sl}/\big( \varepsilon_{k_{F}}^{2}+\delta _{sl}^{2}\big) 
\end{equation}
with $\mu$ in the conduction band;  $\alpha _{xy}$ %the Nernst conductivity is
vanishes  when $\mu$ is in the band gap.  %Equation (14) shows that $\alpha _{xy}$
%the Nernst conductivity exhibits a $1/(E_{F}^{'})^2$ behaviour. 
Thermoelectric transport
%properties 
can be understood by results such as  Eq. (14)  
and agree well with   low-temperature  %experimental 
data \cite%
{PW,YM,KL,AA,JG}  from gapless graphene
in a transverse magnetic field. The Nernst effect discussed here  
exists even without an external magnetic field, being solely driven by the
weak $B$   field and the Berry-curvature. Note that Eq.\ (10) is
more general than Eq.\ (14) since it is valid beyond the $T\to 0$  regime.

The dependence of $\alpha _{xy}$  on the gate voltage (or chemical potential
$\mu$) can be assessed by controlling the band gap, which has been realized
experimentally in graphene \cite{PW,YM,KL,AA,JG}. An enhanced thermoelectric
response is achieved when the bands come close to the Dirac point. In Fig.\
4 we show numerical results for $\alpha_{xy}$, given by Eq.\ (10), as a function of $\mu$ 
at $T=100$ K (left) and $T=200$ K (right). We use $\Delta_{h}=0$ meV and vary the
band gap by off-resonant light such that  $\Delta_{\Omega}=20$ meV (solid), $%
\Delta_{\Omega}=70$ meV (dotted), and $\Delta_{\Omega}=120$ meV
(dot-dashed). We obtain similar results for fixed $\Delta_{\Omega}=0$ meV
and variable $\Delta_{h}$ using values similar  to those of experiments \cite{YZ}. 
%\textbf{
The highest peak value of $\alpha_{xy}$, %thermal conductivity peak (at
near $y\approx 0.4$ %.0 on y-axis)
 in Fig. 4, is 0.4$\, ek_{B}/h \approx$ 52 nA/K. % (solid red line) and similarly can be obtained for other lines.}
Our results show that a certain %significant
thickness or an off-resonant light can significantly affect transport  in TIs at room temperature or even above.
\begin{figure}[ht]
\begin{center}
\vspace*{-0.2cm}
\includegraphics[width=0.48\columnwidth,height=0.40\columnwidth %clip
]{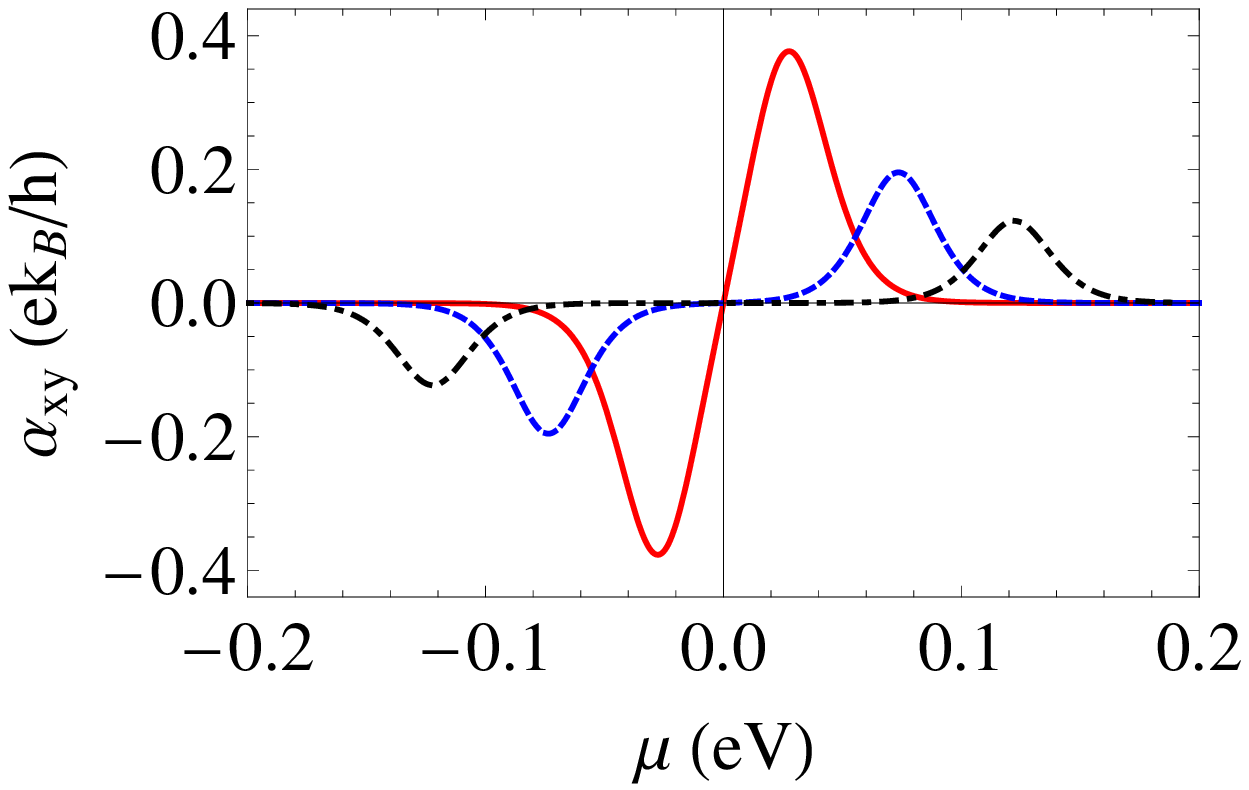}
\includegraphics[width=0.48\columnwidth,height=0.40\columnwidth %clip
]{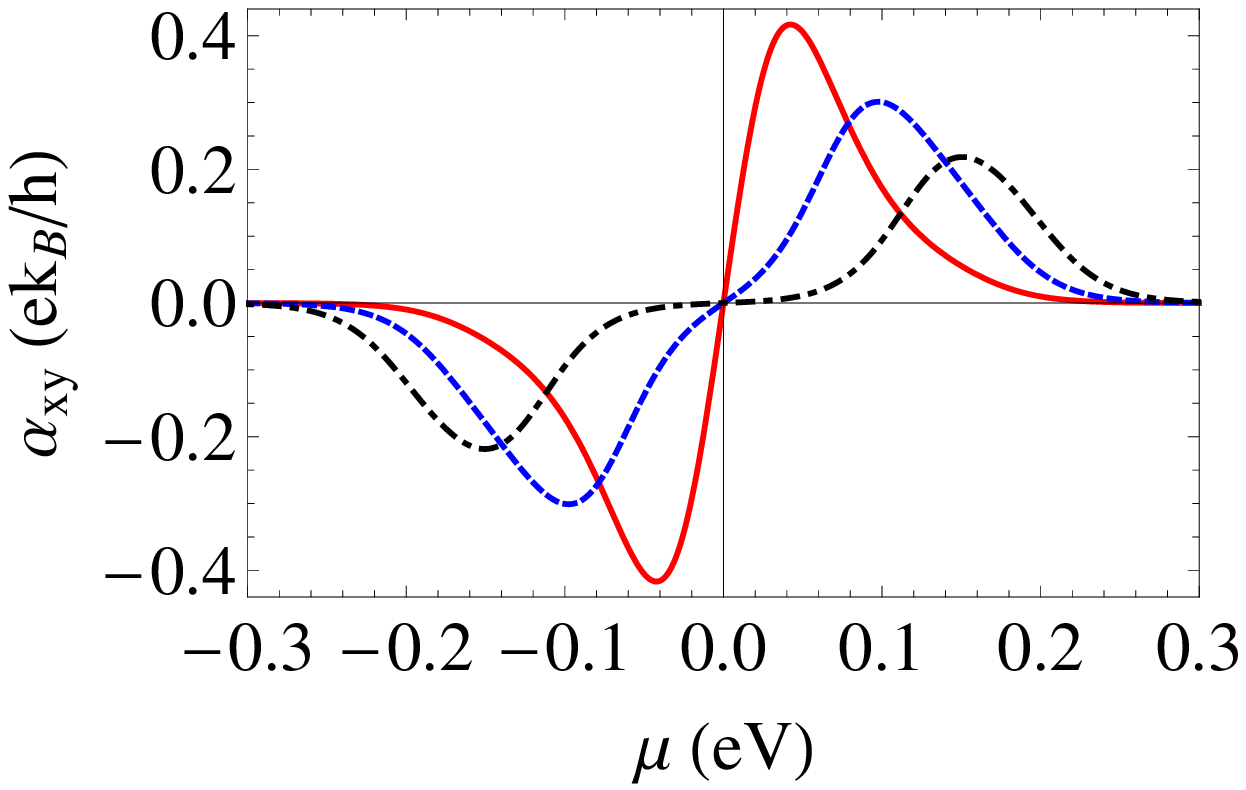}
\end{center}
\vspace{-0.7cm}
\caption{Nernst conductivity vs  $\mu$ for $T$ = 100
K (left) and $T$ = 200 K (right). Here $\Delta_{h}=0$ meV,
$l=+1$, and  $\Delta_{\Omega}= 20$ meV
(red solid), $70$ meV (blue dashed), and $120$ meV
(black dot-dashed). We obtain  similar  results
for $\Delta_{\Omega}$ and $\Delta_{h}$ interchanged.} 
\label{fig:4}
\end{figure}
Figure 5  shows $\alpha_{xy}$ versus $\mu$  for $T=100$ K
(left) and $T=200$ K (right),  $\Delta_{h}=35$ meV, and variable 
$\Delta_{\Omega}=50$ meV (dotted blue), $\Delta_{\Omega}=100$
meV (solid blue). The blue curve is for right polarization of the
light  whereas the black one is for left polarization.
%\textbf{With reference 
Each peak of Fig. 4 is split in two well separated  peaks in both  bands due to the combination of $\Delta_{h} \pm \Delta_{\Omega}$ for $T=100$ K. This is consistent with Figs. 1 and 2. % temperature case.  
As we increase the temperature to 200 K or more, the splitting is suppressed but still persists till room temperature.  We observe shifts of the peaks towards the Dirac point for decreasing $%
\Delta_{\Omega}$, which reflects the reduction of the band gap, and an increase of the
amplitude. %grows with decreasing  gap induced by  $\Delta_{\Omega}$. 
Notice how the sign of $\alpha_{xy}$ is reversed upon reversing the light polarization from right  (blue, $l=+1$) to left (red, $l=-1$). This reversal corresponds to the exchange of the  bands of the symmetric and antisymmetric
surface states shown in Figs. 1 and 2. 
Thus, the transport can be tuned either by  off-resonant
light ($\Delta_{\Omega}$) or by %varying 
the  thickness ($\Delta_{h}$) of the TIs. 
%We also observe  an exchange of the  bands of the symmetric and anti-symmetric
%surface states by reversing the light polarization. 
 This and  the dependence of $\alpha_{xy}$  on the light polarization is, to our knowledge, an entirely new phenomenon. 
\begin{figure}[ht]
\begin{center}
\vspace*{0.1cm}
\includegraphics[width=0.48\columnwidth,height=0.4\columnwidth %clip
]{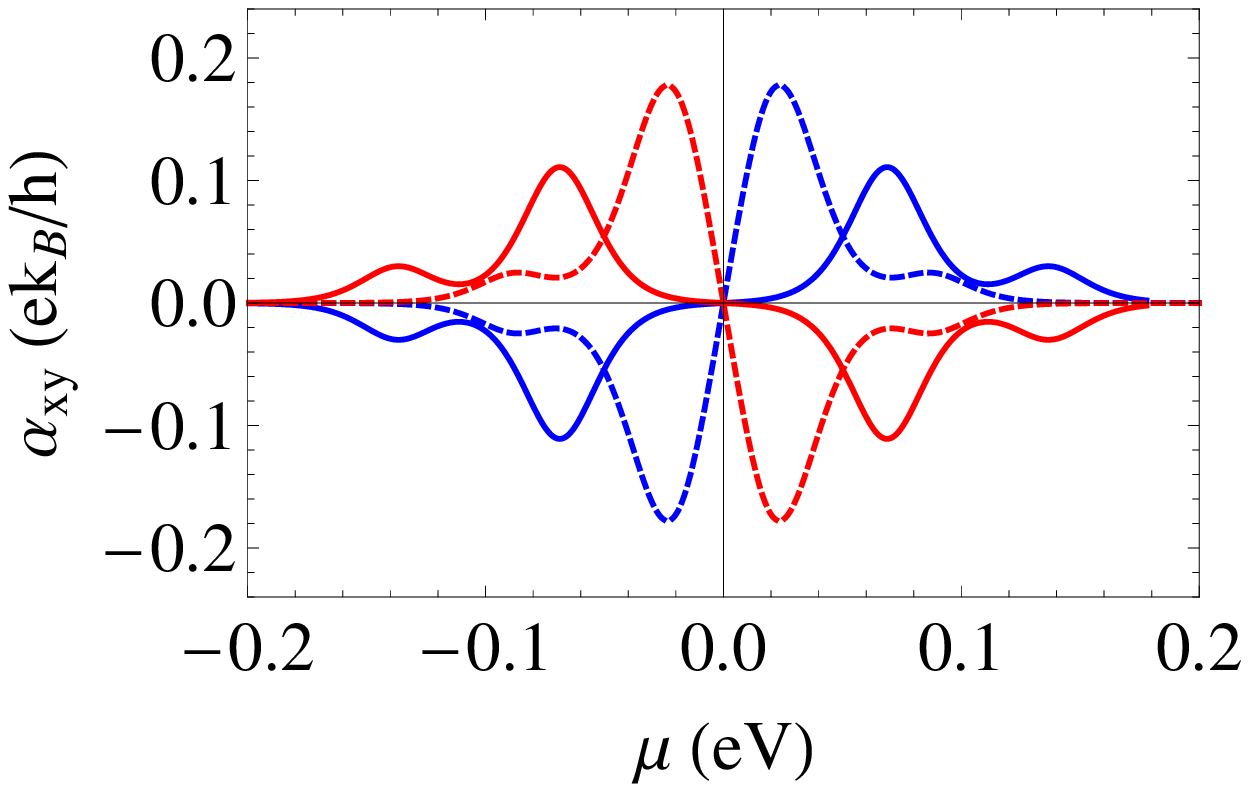} \vspace*{0.4cm}
\vspace*{-0.25cm}
\includegraphics[width=0.48\columnwidth,height=0.4\columnwidth %clip
]{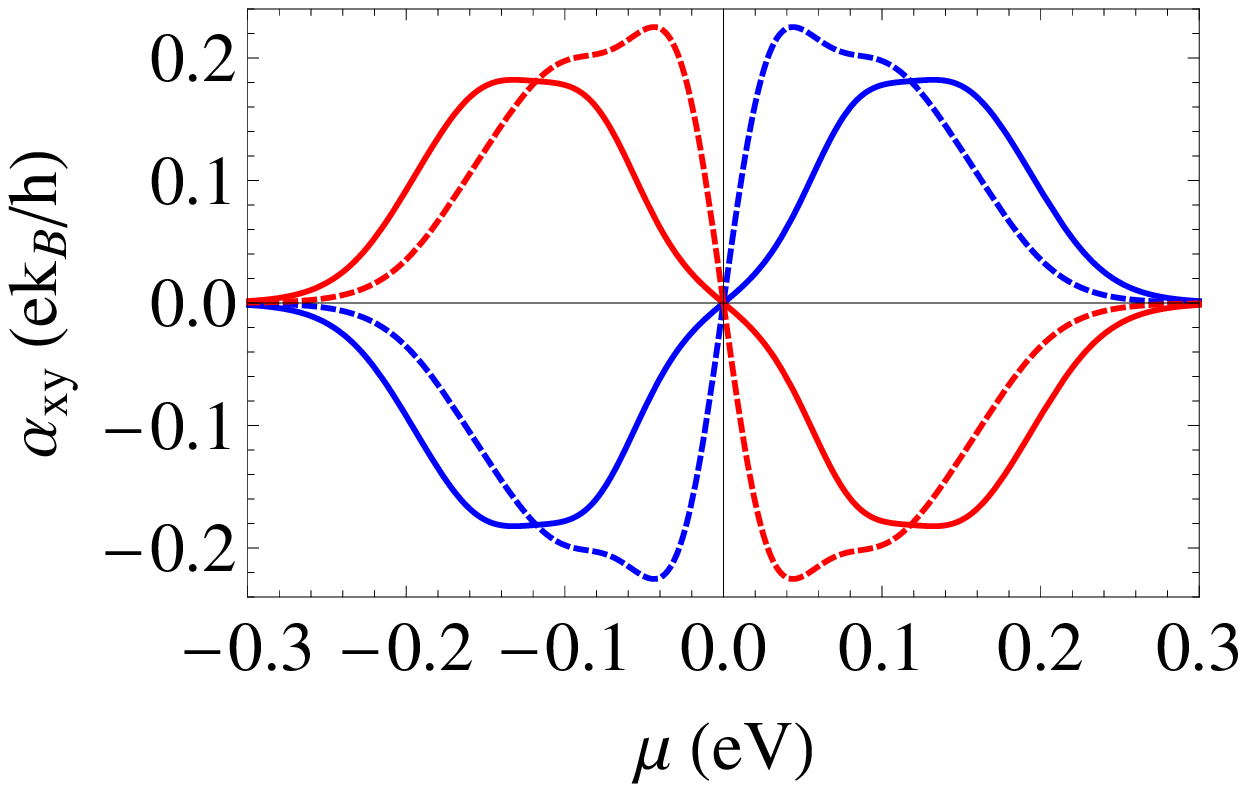}
\end{center}
\vspace{-0.80cm}
\caption{Nernst conductivity vs  $\mu$
for $T$ = 100
K (left) and $T$ = 200 K (right). Here $\Delta_{h}=35$ meV and $\Delta_{\Omega}=100$ meV (solid) and $50$ meV (dotted). Blue and red curves correspond to
right- ($l=+1$) and left- ($l=-1$) %circularly 
polarized light,
respectively.}
\label{fig:5}
\end{figure}
\begin{figure}[bt]
\begin{center}
\vspace{-0.35cm}
\includegraphics[width=0.48\columnwidth,height=0.4\columnwidth %clip
]{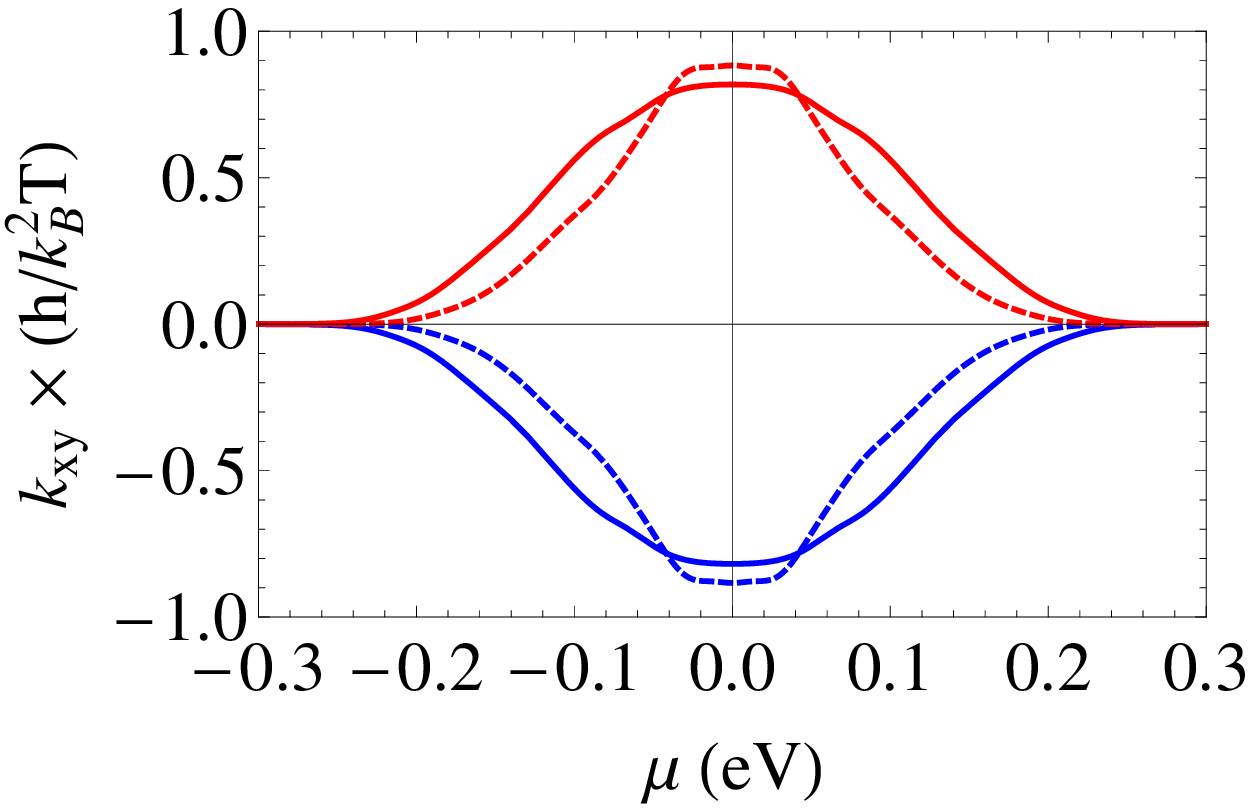} \vspace{-0.15cm}
\hspace{0.2cm}\includegraphics[width=0.48\columnwidth,height=0.4\columnwidth %clip
]{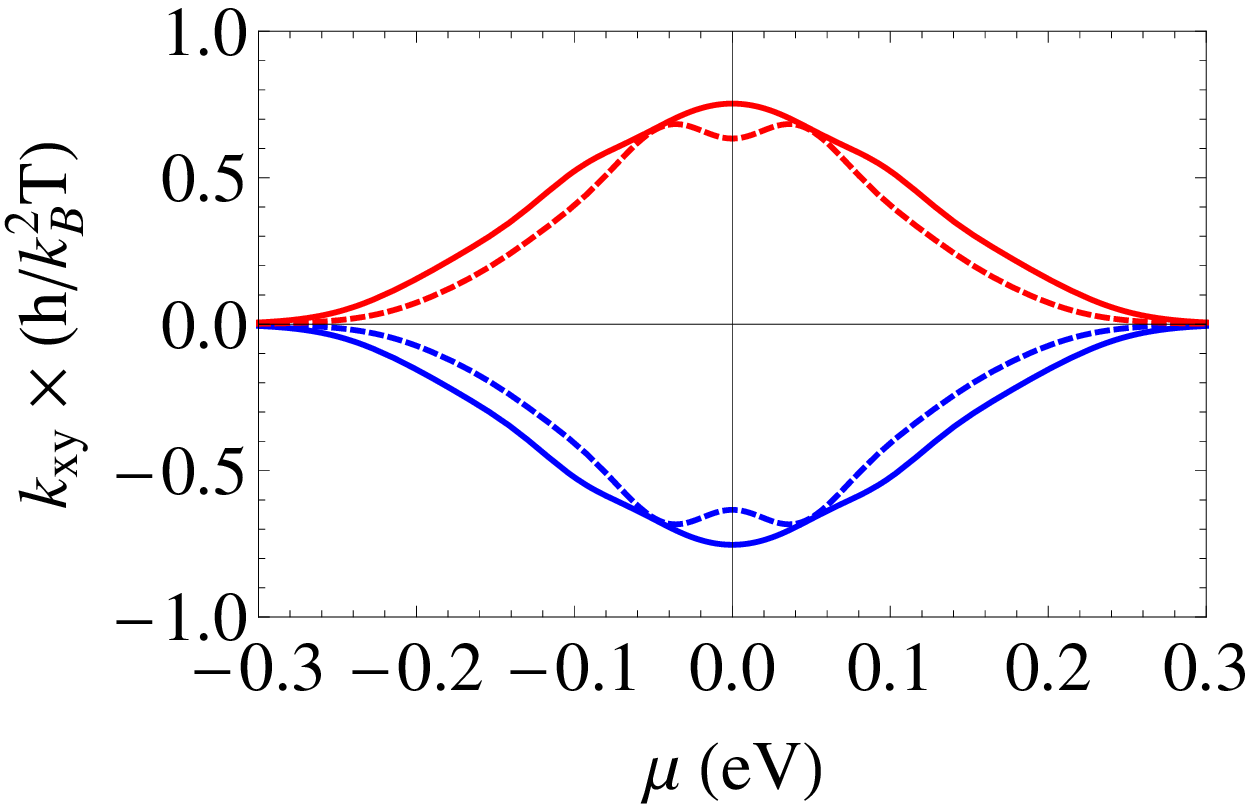}
\end{center}
\vspace{-0.5cm}
\caption{Thermal conductivity vs $\mu$ 
for $T$ = 100
K (left) and $T$ = 200 K (right). Here $\Delta_{h}=35$ meV
and $\Delta_{\Omega}=100$ meV (solid) and $50$ meV (dotted).
Blue and red curves correspond to  right- ($l=+1$) and left-
($l=-1$) %circularly 
polarized light, respectively.}
\label{fig:6}
\end{figure}

In general, it depends on the sign of the Berry-curvature (cf. Fig.\ 1)
whether the Nernst conductivity is positive or negative. Our results are
valid for elevated temperatures in the experimentally relevant range \cite{JG}. 
Moreover, $\alpha_{xy}$ $\neq0$ when $\mu$ is in the band
gap whereas Eq. (14) yields $\alpha_{xy}=0$ since it is the derivative of $%
\sigma _{xy}^{0}$, which is quantized and independent of $\mu$ in this
case.  Notice that given the Mott relations stated above Eq. (12), a similar sign reversal should occur
in the thermal conductivity $\kappa_{xy}$ upon reversing the light polarization. Indeed, Fig. 6, in which we plot  $\kappa_{xy}$ versus $\mu$, obtained numerically from Eq. (11), shows that this is the case: $\kappa_{xy}$ increases linearly with  temperature. In contrast, 
$\sigma _{xy}$, given by Eq. (12), depends very weakly on  temperature.
%\textbf{
The highest peak value of $\kappa_{xy}$, %thermal conductivity peak (at
near $y\approx 1$ %.0 on y-axis)
 in Fig. 6, is $k_{B}^{2}T/h \approx$ 1.2 nAV/K. %}
It is %also 
important to note that there may be additional contributions to  thermoelectric transport properties due to phonons.  However, these contributions are estimated to be   
negligible \cite{TY}  for $T\sim 100$ K. This  tuning of transport  by %applying 
an off-resonant light is pertinent  to thermoelectric device applications. We believe that the 
%thermal Hall conductivity 
$\kappa_{xy}$ and $\alpha_{xy}$ %shown in Fig. 6 
can be measured experimentally in a way similar to that used for bulk ferromagnet
 \cite{WWM}. % and graphene \cite{KL,AA}. %}

All our results, obtained within linear-response theory, rely on the assumption that  the electronic subsystem is not far from thermal equilibrium
when it is exposed to an external off-resonant light. This may not be obvious. However, as argued and explicitly demonstrated in Ref. \onlinecite{TK} by the use of an adiabatic theorem for periodically-driven systems, the transport properties of the nonequilibrium systems are well approximated by those  of the system described by an effective {\it static} Hamiltonian that incorporates  virtual photon absorption processes. Moreover, using the Floquet Fermi golden rule \cite{TK}, it has been demonstrated that excitations in the bands of effective Hamiltonians still require a physical energy greater than the gap. It is in principle possible to absorb energies from photons, but because their frequency is assumed  much larger than the bandwidth, such an absorption %of such large energy 
requires excitations of electrons and many phonons and, therefore, %such a process 
is suppressed. Accordingly, at low temperatures the %property of an "
insulating state of the effective Hamiltonian is protected against electron-phonon interactions by the gap. This holds for  short laser pulses and was fulfilled in recent experiments \cite{YH,HZJ}. Also, due to its topological nature, the effect we obtain should generally be stable against imperfections of the sample.

\textit{Summary}. We  evaluated 
analyticaly and numerically  the band structure and thermoelectric transport properties 
 of ultrathin TIs under the application of \textit{%
off-resonant light}. We showed that by applying a circularly polarized light,
the band gap is tuned and results in  enhanced thermoelectric transport.
Moreover, changing the light polarization from right to left  leads to an exchange  
of the conduction and valence bands of the symmetric and antisymmetric surface states and to a sign reversal of the Nernst and thermal conductivities and of the Berry-curvature-induced orbital magnetization.  
The results present  new opportunities for state-exchanged  excitations under
light and tunable thermoelectric transport properties. 

Our work was supported by  the Canadian NSERC Grant No. OGP0121756. 

$^{\dag}$Electronic address: m.tahir06@alumni.imperial.ac.uk
$^{\star}$Electronic address: p.vasilopoulos@concordia.ca

%\vspace*{1cm}


\begin{thebibliography}{99} 

\bibitem{MZ} M. Z. Hasan and C. L. Kane, Rev. Mod. Phys. \textbf{82} , 3045
(2010); X. L. Qi and S. C. Zhang, {\it ibid} %Rev. Mod. Phys. 
\textbf{83}, 1057 (2011); J. E. Moore, Nature \textbf{464}, 194 (2010). 

\bibitem{HZ} H. Zhang, C.-X. Liu, X.-L. Qi, X. Dai, Z. Fang, and S.-C. Zhang,
Nature Phys. \textbf{5}, 438 (2009).

\bibitem{YZ} Y. Zhang, K. He, C. Z. Chang, C. L. Song, L. L. Wang, X. Chen,
J. F. Jia, Z. Fang, X. Dai, W. Y. Shan, S. Q. Shen, Q. Niu, X. L. Qi, S. C.
Zhang, X. C. Ma, and Q. K. Xue, Nature Phys. \textbf{6}, 584 (2010).

\bibitem{SC} S. Chadov, X.-L. Qi, J. \"{K}ubler, G. H. Fecher, and C. F.
S.-C. Zhang, Nature Material \textbf{9}, 541 (2010).

\bibitem{HL} H. Lin, L. A. Wray, Y. Xia, S. Xu, S. Jia, R. J. Cava, A.
Bansil, and M. Z. Hasan, Nature Material \textbf{9}, 546 (2010).

\bibitem{XZ} X. Zhang, H. Zhang, J. Wang, C. Felser, and S.-C. Zhang, Science 
\textbf{335}, 1464 (2012).

\bibitem{XZJ} X. Zhang, J. Wang, and S.-C. Zhang, Phys. Rev. B \textbf{82},
245107 (2010).

\bibitem{SCN} S. Cho, N.P. Butch, J. Paglione, and M. S. Fuhrer, Nano Lett. 
\textbf{11}, 1925 (2011).

\bibitem{HP} H. Peng, W. Dang, J. Cao, Y. Chen, D. Wu, W. Zheng, H. Li,
Z.-X. Shen, and Z. Liu, Nature Chem. \textbf{4}, 281 (2012).

\bibitem{PG} P. Ghaemi, R. S. K. Mong, and J. E. Moore, Phys. Rev. Lett. 
\textbf{105}, 166603 (2010).

\bibitem{TY} T. Yokoyama and S. Murakami, Phys. Rev. B \textbf{83},
161407(R) (2011).

\bibitem{KU} K. Uchida, S. Takahashi, K. Harii, J. Ieda, W. Koshibae, K.
Ando, S. Maekawa, and E. Saitoh, Nature \textbf{455}, 778 (2008).

\bibitem{DX} D. Xiao, Y. Yao, Z. Fang, and Q. Niu, Phys. Rev. Lett. \textbf{%
97}, 026603 (2006).

\bibitem{DXM} D. Xiao, M. C. Chang, and Q. Niu, Rev. Mod. Phys. \textbf{82},
1959 (2010).

\bibitem{SY} S. Y. Xu, M. Neupane, C. Liu, D. Zhang, A. Richardella, L. A.
Wray, N. Alidoust, M. Leandersson, T. Balasubramanian, J. S. Barriga, O.
Rader, G. Landolt, B. Slomski, J. H. Dil, J. Osterwalder, T. R. Chang, H. T.
Jeng, H. Lin, A. Bansil, N. Samarth, and M. Z. Hasan, Nature Phys. \textbf{8}%
, 616 (2012).

\bibitem{PW} P. Wei, W. Bao, Y. Pu, C. N. Lau, and J. Shi, Phys. Rev. Lett. 
\textbf{102}, 166808 (2009).

\bibitem{YM} Y. M. Zuev, W. Chang, and P. Kim, Phys. Rev. Lett. \textbf{102}%
, 096807 (2009).

\bibitem{KL} K. L. Grosse, M.-H. Bae, F. Lian, E. Pop, and W. P. King,
Nature Nanotechnol.\textbf{\ 6}, 287 (2011).

\bibitem{AA} A. A. Balandin, Nature Material \textbf{10}, 569 (2011).

\bibitem{TK} T. Kitagawa, T. Oka, A. Brataas, L. Fu, and E. Demler, Phys.
Rev. B \textbf{84}, 235108 (2011).

\bibitem{NHL} %\textbf{
N. H. Lindner, G. Refael, and V. Galitski, Nature Physics \textbf{7}, 490 (2011);
N. H. Lindner, D. L. Bergman, G. Refael, and V. Galitski, Phys. Rev. B {\bf 87}, 235131 (2013).

\bibitem{MCR} % \textbf{
M. C. Rechtsman, J. M. Zeuner, Y. Plotnik, Y. Lumer, D. Podolsky, F. Dreisow, S. Nolte, M. Segev, and A. Szameit, Nature \textbf{496}, 196 (2013).

\bibitem{YH} Y. H. Wang, H. Steinberg, P. J. Herrero, and N. Gedik, Science 
\textbf{342}, 453 (2013).

%\bibitem{YO} Y. Onishi, Z. Ren, M. Novak, K. Segawa, Y. Ando, and K. Tanaka,
%arXiv:1403.2492v1.

\bibitem{HZJ} H. Zhang, J. Yao, J. Shao, H. Li, S. Li, D. Bao, C. Wang, and
G. Yang, Sci. Rep. \textbf{4}, 5876 (2014).

\bibitem{OE} D. K. Efimkin and Yu. E. Lozovik, Phys. Rev. B \textbf{87}, 245416 (2013);
Z. Li and J. P. Carbotte, {\it ibid}   \textbf{88}, 045414 (2013);
M. Lasia and L. Brey, {\it ibid}   \textbf{90}, 075417 (2014).

\bibitem{HZL} %\textbf{
H.-Z. Lu, W.-Y. Shan, W. Yao, Q. Niu, and S.-Q. Shen, Phys. Rev. B \textbf{81},
115407 (2010).

\bibitem{JLT} %\textbf{
J. Linder, T. Yokoyama, and A. Sudbo, Phys. Rev. B \textbf{80},
205401 (2009).

\bibitem{CXL} %\textbf{
C.-X. Liu, H. Zhang, B. Yan, X.-L. Qi, T. Frauenheim, X. Dai, Z. Fang, and
S.-C. Zhang, Phys. Rev. B \textbf{81},
041307 (2010).

\bibitem{AG} \'{A}. G\'{o}mez-Le\'{o}n, P. Delplace, and G. Platero, Phys.
Rev. B \textbf{89}, 205408 (2014).

\bibitem{PT} P. Titum, N. H. Lindner, M. C. Rechtsman, and G. Refael,
arXiv:1403.0592v1.

\bibitem{DWQ} D. Xiao, W. Yao, and Q. Niu, Phys. Rev. Lett. \textbf{99},
236809 (2007).

\bibitem{col} L. J. Collins-McIntyre, S. E. Harrison, P. Scho?nherr, N.-J. Steinke, C. J. Kinane,
T. R. Charlton, D. Alba-Veneroa, A. Pushp, A. J. Kellock, S. S. P. Parkin, J. S. Harris, S. Langridge, G. van der Laan, and T. Hesjedal, EuroPhys. Lett., {\bf 107}, 57009 (2014).

\bibitem{gru} M. M. Gruji\'c,  M. Z. Tadi\'c,  and F.  M. Peeters, Phys.
Rev. B \textbf{90}, 205408 (2014).

\bibitem{YK} Y. K. Kato, R. C. Myers, A. C. Gossard, and D. D. Awschalom,
Science \textbf{306}, 1910 (2004).

\bibitem{JP} J. P. Dahlhaus, B. M. Fregoso, and J. E. Moore,
arXiv:1408.6811v1.

\bibitem{JG} J. G. Checkelsky and N. P. Ong, Phys. Rev. B \textbf{80},
081413 (2009).

\bibitem{BSJ} %\textbf{
B. Seradjeh, J. E. Moore, and M. Franz, Phys. Rev. Lett. \textbf{103},
066402 (2009).

\bibitem{WWM}Wei-Li Lee, S. Watauchi, V. L. Miller, R. J. Cava, and N. P. Ong, 
Phys. Rev. Lett. {\bf 93}, 226601 (2004).

\end{thebibliography}
\end{document}